\documentclass[aps,prd,showpacs,amssymb, twocolumn,
nofootinbib] {revtex4}

\begin{document}

\title{Uniqueness of the Fock quantization of the Gowdy $T^3$ model}
\author{ Jer\'onimo Cortez}
\affiliation{Colegio de Ciencia y Tecnolog\'\i a,
Universidad Aut\'onoma de la Ciudad de M\'exico,
Prolongaci\'on San Isidro 151, San Lorenzo Tezonco,
M\'exico D.F. 09790, M\'exico.}
\author{Guillermo A. Mena Marug\'an}\email{mena@iem.cfmac.csic.es}
\affiliation{Instituto de Estructura de la Materia,
CSIC, Serrano 121, 28006 Madrid, Spain.}
\author{Jos\'e M. Velhinho}\email{jvelhi@ubi.pt}
\affiliation{Departamento de F\'{\i}sica, Universidade
da Beira Interior, R. Marqu\^es D'\'Avila e Bolama,
6201-001 Covilh\~a, Portugal.}

\begin{abstract}

After its reduction by a gauge-fixing procedure, the
family of linearly polarized Gowdy $T^3$ cosmologies
admit a scalar field description whose evolution is
governed by a Klein-Gordon type equation in a flat
background in 1+1 dimensions with the spatial topology
of $S^1$, though in the presence of a time-dependent
potential. The model is still subject to a homogeneous
constraint, which generates $S^1$-translations.
Recently, a Fock quantization of this scalar field was
introduced and shown to be unique under the
requirements of unitarity of the dynamics and
invariance under the gauge group of
$S^1$-translations. In this work, we extend and
complete this uniqueness result by considering other
possible scalar field descriptions, resulting from
reasonable field reparameterizations of the induced
metric of the reduced model. In the reduced phase
space, these alternate descriptions can be obtained by
means of a time-dependent scaling of the field, the
inverse scaling of its canonical momentum, and the
possible addition of a time-dependent, linear
contribution of the field to this momentum. Demanding
again unitarity of the field dynamics and invariance
under the gauge group, we prove that the alternate
canonical pairs of fieldlike variables admit a Fock
representation if and only if the scaling of the field
is constant in time. In this case, there exists
essentially a unique Fock representation, provided by
the quantization constructed by Corichi, Cortez, and
Mena Marug\'an. In particular, our analysis shows that
the scalar field description proposed by Pierri does
not admit a Fock quantization with the above unitarity
and invariance properties.

\end{abstract}

\maketitle

\section{Introduction}

The quantization of symmetry reduced models in general
relativity has been intensively studied as a tool to
learn about conceptual and technical issues in quantum
gravity. Specially relevant is the quantization of
gravitational models with local degrees of freedom,
the so called ``midisuperspaces'', since they retain
the field character of general relativity.

In quantum cosmology, the relevance of midisuperspaces
is strengthened by the fact that, in the absence of a
full quantum theory of gravity, their analysis
provides the most solid way to validate or derive a
consistent quantum treatment of the cosmological
inhomogeneities. Remarkably, the only midisuperspace
model whose quantization has been studied with
sufficient detail in cosmology is the family of
linearly polarized Gowdy spacetimes with spatial
topology of a three-torus $T^3$
\cite{misn,berger,hs,guillermo,pierri,ccq-t3,
torre-prd,come,cocome2,cocome,CCMV}.

Gowdy spacetimes are vacuum spacetimes that possess two
spacelike and commuting Killing vectors and whose
spatial sections are compact \cite{gowdy}. Gowdy
proved that any spacetime with these properties must
have spatial sections that are homeomorphic to a
three-torus, a three-handle $S^1\times S^2$, or a
three-sphere $S^3$ (or to a manifold covered by one of
the above). The case of the three-torus is
particularly interesting. All classical solutions to
general relativity {\it start} then in a spacelike
singularity where the area of the two-dimensional
orbits of the Killing isometries vanishes.
Furthermore, this area increases monotonously in the
evolution, so that one can adopt it as time
coordinate. In fact (apart from convenient
normalization factors) this is the standard choice of
time gauge in the description of the Gowdy $T^3$
cosmologies \cite{guillermo}. The condition of linear
polarization, on the other hand, implies that each
Killing vector is hypersurface orthogonal, and
eliminates one of the two local physical degrees of
freedom of the gravitational field.

By means of a dimensional reduction employing one of
the Killing vectors, the family of linearly polarized
Gowdy $T^3$ spacetimes (that we will call Gowdy
cosmologies or Gowdy model from now on) are
classically equivalent to an axisymmetric massless
scalar field propagating on a gravitational background
in 2+1 dimensions. Then, a quantization of this scalar
field provides essentially a quantum theory for the
Gowdy model. This was the procedure followed by Pierri
to construct a Fock quantization of the Gowdy
cosmologies \cite{pierri}. However, it was soon
pointed out that this quantization is not fully
satisfactory: the classical evolution of the scalar
field cannot be implemented as a quantum unitary
transformation \cite{ccq-t3,torre-prd,come}. To
recover a unitary dynamics, a different choice of
``fundamental field'' for the Gowdy model and a
suitable Fock quantization of it was recently proposed
\cite{cocome2,cocome}. Actually, this new choice of
the scalar field is the result of a different
parameterization of the metric of the Gowdy
cosmologies \cite{cocome}. In the following, we will
refer to the field parameterization of the metric,
Fock representation and quantization of the reduced
Gowdy model introduced in Refs. \cite{cocome2,cocome}
and later elaborated in Ref. \cite{CCMV} by Corichi,
Cortez, Mena Marug\'an, and Velhinho as the CCMV ones.

In the quantization process that leads to the CCMV
representation for the Gowdy model, there are three
steps where one performs choices whose modification
might result in an inequivalent Fock quantization
\cite{cocome}. First, there is the choice of gauge
that allows to eliminate most of the constraints of
the model and reduce the system. Second, one makes a
choice of parameterization for the reduced Gowdy
metric that determines which scalar field is
considered as fundamental. Finally, one has to make a
choice of quantum representation for this scalar
field, as systems with fieldlike degrees of freedom
generally admit inequivalent quantizations. For Fock
representations, this amounts to an ambiguity in the
selection of the ``one-particle'' Hilbert space, which
is fixed by a choice of complex structure (see e.g.
Refs. \cite{ash-mag,poincare,wald}). The aim of the
present work is to demonstrate that, if the choice of
gauge for the Gowdy cosmologies is fixed, the CCMV
quantization for the resulting reduced model is
essentially unique under a set of natural
requirements. This will complete previous work
presented in Ref. \cite{CCMV} which already proves the
uniqueness of the Fock representation with respect to
the choice of complex structure. We will extend that
analysis to take into account different
parameterizations of the reduced Gowdy metric which
select distinct scalar fields as the fundamental
object to be quantized.

Several reasons justify the importance of this result.
On the one hand, the obtained uniqueness guarantees
that the physics of the quantum cosmological model
does not depend on the choice of parameterization or
the particular Fock representation selected, providing
significance to the predictions. On the other hand,
even if one adopted a distinct kind of quantization,
not necessarily equivalent to a Fock one, like e.g. a
polymerlike quantization \cite{polymer}, there ought
to exist a regime in which a Fock quantum theory were
recovered. This condition can hardly be used to
control the acceptable quantizations of the system
unless one can specify such a Fock representation.
Finally, the result has a conceptual interest by
itself, since it shows that it is possible to attain
uniqueness even for non-stationary systems and in the
framework of standard quantum field theory.

The paper is organized as follows. In Sec.
\ref{context} we first pose the problem, discussing
the freedom available in the choice of the scalar
field as the basic fieldlike variable for the reduced
Gowdy model. Under reasonable demands, this freedom
consists just in time-dependent canonical
transformations in the reduced phase space that scale
the field by a positive function of time and its
momentum by the inverse factor. In addition, the
momentum is allowed to get a linear contribution of
the field, with a time dependent coefficient. Sec.
\ref{themodel} reviews the CCMV quantization of the
reduced Gowdy model. We then analyze in Secs.
\ref{alternate} and \ref{compleuni} the alternate Fock
quantizations obtained by adopting different choices
of fundamental field. In addition to a unitary
implementation of the dynamics, we demand that these
representations satisfy a natural condition concerning
the only remaining constraint of the reduced model.
Namely, we require invariance under the corresponding
gauge group. The proof that all such quantizations are
equivalent to the CCMV one is presented in Secs.
\ref{nogo} and \ref{equivrepre}. Finally we conclude
and summarize our results in Sec. \ref{summary}. Two
appendices are added. In Appendix \ref{proof} we
explain some calculations employed in our uniqueness
proof. Appendix \ref{vacham} proposes a criterion to
fix the linear contribution of the field to the
momentum.

\section{The context}\label{context}

Let us start with the metric of the Gowdy spacetimes
in coordinate systems adapted to the two axial Killing
vector fields, so that these are identified as
$\partial_{\sigma}$ and $\partial_{\delta}$ for
certain coordinates $\sigma, \delta \in S^1$. Given
the hypersurface orthogonality of these Killing
vectors, the induced metric can be parameterized in
terms of three fields that depend only on the time
coordinate $t>0$ and one spatial coordinate $\theta\in
S^1$ \cite{come}. These fields describe the norm of
one of the Killing vectors (e.g. $\partial_{\delta}$),
the area of the orbits of the group of isometries, and
the scale factor of the metric induced on the set of
group orbits. For instance, the parameterization used
by CCMV in Ref. \cite{cocome} is
\begin{eqnarray} \label{metricunred}
ds^2&=&e^{\bar{\gamma}
-(\xi/\sqrt{\tau})-\xi^{2}/(4\tau)}\left(-\tau^2
{N_{_{_{\!\!\!\!\!\!\sim}}\;}}^{2} dt^2+[d\theta +
N^{\theta}dt]^2\right) \nonumber\\
&+& \tau^2 e^{-\xi/\sqrt{\tau}}d\sigma^2 +
e^{\xi/\sqrt{\tau}}d\delta^2,
\end{eqnarray} where ${N_{_{_{\!\!\!\!\!\!\sim}}\;}}$
is the densitized lapse function, $N^{\theta}$ the
non-vanishing component of the shift vector, and the
three fields that parameterize the metric are $\xi$,
$\tau>0$, and $\bar{\gamma}$.

The model is subject to the $\theta$-diffeomorphisms
and Hamiltonian constraints. The standard gauge fixing
for the $\theta$-diffeomorphisms imposes the
homogeneity of the phase space variable that generates
conformal transformations of the metric induced on the
set of group orbits. Actually, this condition fixes
only the inhomogeneous part of the
$\theta$-diffeomorphisms constraint. The homogeneous
part, $C_0$, which generates $S^1$-translations,
remains as a constraint on the system. After a partial
reduction, the phase space variable used in our
gauge-fixing condition and its canonically conjugate
variable are determined except for their zero modes.
These zero modes are described by a pair of
canonically conjugate homogenous variables $(Q,P)$
\cite{come,cocome}. Finally, the system is
deparameterized by choosing as time coordinate the
area of the orbits of the isometry group, apart from a
proportionality factor. The subsequent reduction leads
to a system whose degrees of freedom correspond just
to one scalar field plus the ``point-particle''
canonical pair $(Q,P)$, and that is subject to the
constraint $C_0$. With a convenient selection of the
proportionality factor in our choice of time, the
dynamics of the field sector can be decoupled from the
homogenous pair $(Q,P)$. These two homogenous
variables are in fact constants of motion. Finally,
both with the CCMV parameterization of the metric or
with the one adopted in Refs. \cite{pierri,come}, the
field dynamics is given by a Klein-Gordon type
equation that is invariant under $S^1$-translations,
thought explicitly time dependent.

Let us describe  the reduced model in more detail,
e.g. for the CCMV parameterization
(\ref{metricunred}). The gauge-fixing conditions are
then $P_{\bar{\gamma}}=\oint d\theta
P_{\bar{\gamma}}/(2\pi):=-e^P$ (restricted to
solutions with $P\in \mathbb{R}$) and $\tau=t e^P$
(time gauge). Here, $P_{\bar{\gamma}}$ is the momentum
canonically conjugate to $\bar{\gamma}$. The reduced
metric, expressed in the CCMV parameterization, is
obtained from Eq. (\ref{metricunred}) with $\tau=t
e^P$ and \cite{cocome}
\begin{equation}
2\pi \bar{\gamma} e^P=-Q-i\sum_{n=-\infty, n \neq
0}^{\infty}\oint d\bar{\theta}\frac{e^{i n
(\theta-\bar{\theta})}}{n}P_{\xi} \xi^{\prime}+
tH,\end{equation} where $P_{\xi}$ is the canonical
momentum of the remaining field $\xi$, the prime
stands for the derivative with respect to $\theta$,
and $H$ is the (reduced) Hamiltonian that generates
the evolution \cite{note0}:
\begin{equation}\label{hami}
H=\frac{1}{2}\oint d \theta
\left[P^{2}_{\xi}+(\xi^{\prime})^{2}+
\frac{1}{4t^{2}}\xi^{2}\right].
\end{equation}
The associated field equation is
\begin{equation} \label{KG}
\ddot{\xi}-\xi^{\prime\prime}+\frac{\xi}{4t^{2}}=0,
\end{equation}
with the derivative with respect to $t$ denoted by a
dot. Finally, the only constraint of the reduced
system is
\begin{equation} \label{oconst} C_0=\frac{1}{\sqrt{2\pi}}
\oint d\theta P_{\xi} \xi^{\prime} =0.\end{equation}

Of course, when parameterizing the reduced metric in
terms of a scalar field and taking this object as the
variable to be quantized, one is introducing a choice
of (part of the) set of basic variables. However, the
scalar field parameterization of the reduced metric of
the Gowdy model is certainly not unique. In this
sense, there is no scalar field theory canonically
associated with the gauge-fixed Gowdy model, but
rather an infinity of them.

Nonetheless, it is most reasonable to consider only
field parameterizations satisfying certain amenable
properties. We discuss now the class of field
parameterizations analyzed in this work, determined by
a set of natural requirements. For definiteness, we
take the CCMV parameterization as the reference one,
and express alternate parameterizations in terms of
it. First, we consider exclusively scalar fields which
provide a local and (explicitly)
coordinate-independent parameterization of the norm of
the Killing vector $\partial_{\delta}$ on each section
of constant time in the reduced model [possibly
together with the variables that describe the
point-particle degree of freedom]. In this way, the
allowed field reparameterizations are local and
commute both with the isometry group and with the
gauge group of translations in $\theta\in S^1$. In
particular, this guarantees that the corresponding
field dynamics is local and $\theta$-independent, so
that the invariance under $S^1$-translations is
preserved. Besides, the second-order field equation
should be kept linear and homogenous, so that the
space of solutions remains a linear space. Finally, it
is convenient to preserve the decoupling between the
fieldlike and point-particle degrees of freedom (see
however our comments below). With these premises, the
possible field redefinitions in the reduced Gowdy
model consist in scalings of the field $\xi$ by a
function depending exclusively on time \cite{noteZ2}.

Therefore, from now on we will concentrate our
discussion on time-dependent scalings of the field in
the reduced model. This type of scalings can always be
completed into a time-dependent canonical
transformation in the reduced phase space. The
canonical momentum of the field suffers the inverse
scaling. We will also allow for a linear contribution
of the field to the new momentum, with a
time-dependent coefficient. This contribution is
local, preserves the decoupling with the
point-particle degrees of freedom, and is compatible
with all linear structures on phase space, as well as
with $S^1$-translation invariance.

It is not difficult to check that the process of first
fixing the gauge and then performing one of the above
time-dependent canonical transformations is equivalent
to carry out first a time-independent canonical
transformation in the unreduced phase space (with the
role of time coordinate played by the corresponding
internal time variable) and afterwards the gauge
fixing. One could further ask whether these canonical
transformations in the unreduced phase space
correspond just to field reparameterizations of the
unreduced metric (without including the momenta). This
will be the case only if the transformation is a
contact one in the unreduced configuration space of
metric fields. However, taking Eq. (\ref{metricunred})
as reference, we  see that scalings by functions $F$
of $\tau e^{-P}$ (the internal time) will depend on
the momentum variable $P$ unless the scaling is
trivial. Nonetheless, notice that the Klein-Gordon
equation (\ref{KG}) obtained after reduction is not
modified if the field is multiplied by a function of
$P$, which is a constant of motion. If we allow for
this kind of multiplication, the change of field can
be regarded as a contact transformation in the
unreduced configuration space if and only if there
exists a function $L$ such that $L(P) F(\tau e^{-P})$
is independent of $P$. This happens only if
$F(\tau)=\tau^a$ for a certain power $a\in \mathbb{R}$
[then $L(P)=e^{aP}$]. In fact, this occurs in the case
of the field parameterization employed in Ref.
\cite{pierri}, which can be obtained with
$F(\tau)=1/\sqrt{\tau}$ (see Appendix A in Ref.
\cite{cocome}). Multiplication by the function $L(P)$
leaves, nevertheless, a trace in the reduced
Hamiltonian [see e.g. Eq. (\ref{hami})], so that it
actually couples the dynamics of the fieldlike and the
point-particle degrees of freedom of the reduced model
\cite{note}.

Summarizing, for canonical transformations in the
reduced phase space that scale the field by a power of
the time coordinate, and only for them, the
transformation can be understood as the result of a
change of field parameterization of the unreduced
metric followed, after reduction, by multiplication by
a constant of motion in order to decouple the
fieldlike and the point-particle degrees of freedom.
We will however maintain the generality of our
analysis and consider all reasonable scalar field
parameterizations of the reduced metric, so that we
will not restrict our discussion to this specific
subfamily of scalings. In fact, we will see that our
results do not depend on whether one imposes or not
this restriction.

Owing to the time dependence of the considered
canonical transformations in the reduced phase space,
the choice of fundamental scalar field can have a
large impact on the quantization of the reduced Gowdy
model. In fact, since two candidate fields are related
by a time-dependent scaling, the evolution of both
sets of variables is effectively different. It may
then happen that, upon quantization, the dynamics of
one of the fields admits a unitary implementation,
whereas the dynamics of the other does not. Note also
that, if one declares a certain field description to
be fundamental, a quantization based on another field
(related to the first one by a time-dependent
transformation) can be seen as a quantization of the
fundamental field using seemingly awkward
time-dependent variables, instead of the natural field
variables. However, in the context of the Gowdy model
there is {\it a priori} an inherent freedom to choose
the field parameterization of the reduced metric.
Thus, any proposal to single out a field
parameterization should be based on criteria such as
the feasibility of the quantization and its
consistency.

Given the central role that the unitarity of the
evolution plays in the quantum theory (particularly
within the Hilbert space approach), it is certainly
desirable that the selected field parameterization
allows, upon quantization, a unitary implementation of
the classical evolution of the scalar field. As we
have commented, the CCMV formulation admits a Fock
quantization that satisfies this condition. Besides,
the remaining constraint in the scalar field theory
$C_0$ is naturally quantized. So, the outcome of Refs.
\cite{cocome2,cocome} is a consistent, rigorous
quantization of the gauge-fixed Gowdy cosmologies with
unitary evolution.

An important issue, related to the question of
unitarity of the dynamics, is the uniqueness of the
quantum theory. For the reduced Gowdy model with the
CCMV parameterization, it has been demonstrated that
the proposed Fock quantization is indeed unique, under
the following conditions on the quantum representation
of the scalar field \cite{CCMV}. First, one demands a
unitary implementation of the classical evolution.
Second, one asks for  a natural invariant
implementation of the constraint $C_0$, in the sense
that the Fock state --or the complex structure-- that
defines the field representation is required to be
invariant under the gauge group of $S^1$-translations
generated by the constraint \cite{Z5}. The CCMV
representation satisfies these conditions and it turns
out that any representation which does so is unitarily
equivalent to it. Thus, as long as the field
parameterization of the reduced Gowdy model is fixed
(and the invariance condition is fulfilled), the
requirement of unitary dynamics selects a unique Fock
quantization.

In the present work we will considerably deepen this
uniqueness result by showing that it is maintained
when the alternate scalar field parameterizations of
the reduced Gowdy model discussed above are allowed.
In principle, it might happen that a unitary dynamics
could be achieved in a certain Fock quantization of
some different field description, and that the new
quantum theory be physically distinct from the CCMV
one. For instance, this would occur if the quantum
operators corresponding to the CCMV scalar field in
the new description failed to define a representation
equivalent to that introduced in Refs.
\cite{cocome2,cocome}. We will show that this is not
the case: for any  scalar field parameterization, if a
Fock representation exists satisfying the unitary
implementability of the corresponding dynamics and the
invariance under the gauge group of
$S^1$-translations, it is guaranteed that the
evolution of the CCMV field is well defined and
unitary in the new description. By the results of Ref.
\cite{CCMV}, the representation is then the same as
that of Refs. \cite{cocome2,cocome} (modulo unitary
equivalence).

\section{CCMV quantization}
\label{themodel}

We will now briefly review the scalar field
formulation of the reduced Gowdy model obtained in
Refs. \cite{cocome2,cocome} and its proposed
quantization. We obviate the point-particle degrees of
freedom because, being finite in number, they play no
role in the discussion of the uniqueness of the
quantization. For the same reason, we also obviate the
homogeneous mode of the field (see below).

We remember that the fieldlike degrees of freedom of
the reduced model are described in the CCMV
parameterization by the field $\xi$ and its momentum
$P_{\xi}$. Its dynamics is governed by the
time-dependent Hamiltonian (\ref{hami}), which is
invariant under the group of $S^1$-translations:
\begin{equation} \label{o4}
T_{\alpha}: \theta\mapsto\theta+\alpha
\quad\quad \forall\alpha\in S^1.
\end{equation}
These translations are gauge symmetries of the reduced
model, generated by the only constraint that remains
on the system, namely $C_0$.

Taking into account that the canonical fields $\xi$
and $P_{\xi}$ are periodic in $\theta$, we can expand
them in Fourier series:
\begin{eqnarray} \xi(\theta,t)&=&
\sum_{n=-\infty}^{\infty}
\xi_{n}(t)\frac{e^{in\theta}}{\sqrt{2\pi}}, \nonumber
\\\label{o5}
P_{\xi}(\theta,t)&=&\sum_{n=-\infty}^{\infty}
P^{n}_{\xi}(t)\frac{e^{in\theta}}{\sqrt{2\pi}}.
\end{eqnarray}
The Fourier coefficients $\xi_n(t)$ and
$P_{\xi}^{-n}(t)$ are canonically conjugate pairs of
variables, which alternatively describe the degrees of
freedom of the system. Since the field $\xi$ and its
momentum are real, the Fourier coefficients satisfy
the reality conditions $\xi_n^{*}(t)=\xi_{-n}(t)$ and
$[P_{\xi}^{n}(t)]^{*}=P_{\xi}^{-n}(t)$. Here, the
symbol $*$ denotes complex conjugation.

As we mentioned, we will  obviate the zero modes of
these fields for simplicity. To describe all other
modes we introduce the set of variables
\begin{eqnarray}
b_m(t)&:=&\frac{m\xi_{m}(t)+iP_{\xi}^{m}(t)}
{\sqrt{2m}},\nonumber \\ \label{o6}
b_{-m}^*(t)&:=&\frac{m\xi_{m}(t)-iP_{\xi}^{m}(t)}
{\sqrt{2m}},\end{eqnarray} together with their
respective complex conjugate $b_{m}^{*}(t)$ and
$b_{-m}(t)$, where $m\in \mathbb{N}$ is any strictly
positive integer. Besides, we will assemble them in
the column vectors
\begin{equation}
\label{obcal}B_m(t):=\left( b_m(t), b_{-m}^*(t),
b_{-m}(t), b_{m}^*(t)\right)^{T}.\end{equation}
The symbol $T$ denotes the transpose.

The variables $\{B_m(t)\}$ simply acquire a phase
under the action of translations $T_{\alpha}$: \[
b_{\pm m}(t)\mapsto e^{\pm im\alpha}b_{\pm m}(t),
\quad \quad b_{\mp m}^{*}(t)\mapsto e^{\pm
im\alpha}b_{\mp m}^{*}(t).
\]

The classical evolution of the system is expressed in
terms of these variables as follows \cite{cocome2}.
Evolution from data $\{B_m(t_0)\}$ at a certain
instant of time $t_0$ to $\{B_m(t)\}$ at a different
time $t$ is given by a classical evolution operator
$U(t,t_0)$, which,  for these variables, takes  the
block-diagonal form
\begin{eqnarray} B_m(t)&=&U_m(t,t_0)
B_m(t_0),\nonumber \\  U_m(t,t_0)&=&
W(x_m)W(x^0_m)^{-1},\label{o10}
\end{eqnarray} with $x_m:=mt$, $x^0_m:=mt_0$, and
\[ W(x)= \left(
\begin{array}{cc}
{\cal W}(x) & {\bf 0}  \\
{\bf 0} & {\cal W}(x)
\end{array} \right),\quad {\cal W}(x)= \left(
\begin{array}{cc}
c(x) & d(x) \\
d^*(x) & c^*(x)
\end{array} \right),\]
\begin{eqnarray}
d(x)&:=&\sqrt{\frac{\pi x}{8}}\left[
\left(1+\frac{i}{2x}\right)H_0^*(x)-iH_1^*(x)\right],
\nonumber\\
c(x)&:=&\sqrt{\frac{\pi x}{
2}}H_0(x)-d^{*}(x).\label{cdfun}
\end{eqnarray} Here, ${\bf 0}$ is
the zero $2\times 2$ matrix and $H_j$ ($j=0,1$) is the
$j$-th order Hankel function of the second kind
\cite{abra}. Since $|c(x)|^2 - |d(x)|^2=1$, the map
defined by $U_m(t,t_0)$ is a Bogoliubov
transformation.

One can check that the evolution matrices $U_m(t,t_0)$
are then block-diagonal in $2\times 2$ blocks [as
$W(x)$ above], with the two diagonal blocks being
equal to the same $2\times 2$ matrix ${\cal
U}_m(t,t_0)$:
\begin{eqnarray} {\cal U}_m(t,t_0)
&:=&\left( \begin{array}{cc}\alpha_m(t,t_0) &
\beta_m(t,t_0)
\\ \beta^*_m(t,t_0) & \alpha^*_m(t,t_0)
\end{array} \right),\nonumber\\
\alpha_m(t,t_0)&:=&c(x_m)c^*(x^0_m)-d(x_m)d^*(x^0_m),
\nonumber\\
\label{o15} \beta_m(t,t_0)&:=&d(x_m)c(x^0_m)-c(x_m)
d(x^0_m).
\end{eqnarray}
For further calculations we also note that, from Eq.
(\ref{cdfun}) and the asymptotic behavior of the
Hankel functions \cite{abra}, the functions $d(x)$ and
$c(x)-e^{i\pi/4}e^{-ix}$ tend to zero in the limit
$x\to\infty$. In particular, it then follows that, for
every fixed $t_0$ and $t$, the sequences
$\{\beta_m(t,t_0)\}$ and
$\{\alpha_m(t,t_0)-e^{-im(t-t_0)}\}$ vanish in the
limit $m\to\infty$.

The CCMV quantization of the reduced Gowdy model is
defined by using a representation for $\xi$ on a
fiducial Fock space which allows a unitary
implementation of the dynamics as well as of the group
of $S^1$-translations \cite{cocome2,cocome}. This
quantization is of the Fock type, i.e. it is defined
by a Hilbert space structure in phase space (or in the
space of smooth solutions), which in turn is uniquely
determined by a complex structure. The resulting
Hilbert space is the so-called one-particle space,
from which the quantum Fock space is constructed (see
e.g. \cite{poincare,fock}).

The procedure to introduce this quantization is the
following. We first fix, once and for all, a reference
time $t_0$ and identify the phase space as the space
of Cauchy data at $t=t_0$, expressed e.g. by the
linear combinations of Fourier components
$\{B_m(t_0)\}$ defined by Eqs. (\ref{o6},\ref{obcal})
for $t=t_0$. In order to simplify the notation, we
will denote $\{B_m(t_0)\}$ simply as $\{B_m\}$ from
now on, understanding the evaluation at the reference
time $t_0$. Thus, the fields at the instant $t_0$ play
in our case the same role as the time-zero fields in
standard quantum field theory in Minkowski spacetime.
However, owing to the compactness of the spatial
manifold $S^1$, the quantum counterparts of the
Fourier components need not be smeared in Fourier
space, i.e. one obtains well defined operators
$\hat{\xi}_n(t_0)$ and $\hat{P}_{\xi}^n(t_0)$
(satisfying the reality conditions), as well as
$\hat{b}_m$, $\hat{b}_{-m}$ and their corresponding
adjoints (like for their classical counterpart
$\{B_m\}$, evaluation at $t=t_0$ is implicitly
understood for these operators in the following). The
two sets of operators are of course related like in
Eq. (\ref{o6}) for $t=t_0$. On the other hand,
operators like $\hat\xi(\theta,t_0)$ remain formal,
with a well defined meaning assigned only to
appropriately smeared fields.

The complex structure $J_0$ selected in Refs.
\cite{cocome2,cocome} to carry out the quantization
takes the form of a block-diagonal matrix in the basis
$\{B_m\}$, with $4\times 4$ blocks $(J_0)_m={\rm
diag}(i,-i,i,-i)$. With this choice of complex
structure, the variables $\{B_m\}$ are quantized as
the annihilation and creation operators of the Fock
representation. Then the corresponding Fock vacuum is
char\-acterized by the conditions
$\hat{b}_m|0\rangle=\hat{b}_{-m}|0\rangle=0$ $\forall
m\in \mathbb{N}$ \cite{noteZ1}. Besides, the
invariance of the complex structure $J_0$ under the
group of $S^1$-translations (\ref{o4}) guarantees an
invariant unitary implementation for this gauge group.
One thus obtains unitary operators $\hat{T}_{\alpha}$
$\forall \alpha\in S^1$ which leave the vacuum
invariant and satisfy $\hat{T}_{\alpha}^{-1}
\hat{b}_m\hat{T}_{\alpha}= e^{im\alpha}\hat{b}_m$
(with similar actions on $\hat{b}_{-m}$ and
$\hat{b}_{\pm m}^{*}$). Therefore the vacuum is
annihilated by the generator of the unitary group
$\hat{T}_{\alpha}$, which is the quantum constraint
operator.

Most importantly, the classical dynamics turns out to
admit also a unitary implementation in this Fock
representation \cite{cocome2,cocome}. At this stage,
it is worth recalling that, given a Fock space defined
by a complex structure $J$, one can obtain on it a
unitary implementation of a symplectic transformation
$A$ if and only if its antilinear part $A_J=(A+JAJ)/2$
is Hilbert-Schmidt on the one-particle Hilbert space
(determined by $J$) or, equivalently, if and only if
$J-AJA^{-1}$ is Hilbert-Schmidt \cite{un1}. In the
case of the family of symplectic transformations
defined by the classical dynamics, namely the linear
transformations $U(t,t_0)$ for each $t$, and
considering the CCMV representation for the reduced
Gowdy model, the Hilbert-Schmidt condition for the
existence of a unitary implementation amounts to
demanding that $\sum_{m=1}^{\infty}|\beta_m(t,t_0)|^2$
be finite $\forall t>0$. This square summability
condition is indeed satisfied, as shown in Refs.
\cite{cocome2,cocome}. Hence, there exist unitary
operators $\hat{ U}(t,t_0)$ such that, $\forall m\in
\mathbb{N}$,
\begin{eqnarray} \label{o19} \hat{U}^{-1}(t,t_0)
\hat{b}_m\hat{ U}(t,t_0)&=&
\alpha_m(t,t_0)\hat{b}_m+\beta_m(t,t_0)
\hat{b}^{\dag}_{-m},\nonumber\\
\label{z1} \hat{U}^{-1}(t,t_0)
\hat{b}^{\dag}_{-m}\hat{ U}(t,t_0)&=&
\beta_m^*(t,t_0)\hat{b}_m+\alpha_m^*(t,t_0)
\hat{b}^{\dag}_{-m}.\nonumber\end{eqnarray}

To conclude, let us also remind that two complex
structures $J$ and $J'$ give rise to unitarily
equivalent Fock representations if and only if $J-J'$
is a Hilbert-Schmidt operator on the one-particle
Hilbert space defined by $J$ (or $J'$) (see e.g. Refs.
\cite{ash-mag,ash-mag2}). We will say that such
complex structures are equivalent, $J\sim J'$.
Therefore, a symplectic transformation $A$ is
unitarily implementable on a Fock space defined by $J$
if and only if the complex structures $J$ and
$AJA^{-1}$ are equivalent.

\section{Alternate field formulations}
\label{alternate}

We will now start to investigate alternate
quantizations of the reduced Gowdy model derived from
other reasonable field parameterizations of the
reduced metric. According to our discussion in Sec.
\ref{context}, we consider different scalar field
formulations that are obtained by a time-dependent
scaling of the CCMV field $\xi$. In the reduced phase
space, the reformulation of the model is provided by a
time-dependent canonical transformation of the type:
\begin{equation}
\label{1} \varphi:=F(t) \xi, \quad
P_{\varphi}:=\frac{P_{\xi}}{F(t)}+G(t)\xi ,
\end{equation}
where $(\varphi,P_{\varphi})$ are the new canonical
fieldlike variables (to be considered now as
fundamental), and $F(t)$ and $G(t)$ are real
continuous functions on $\mathbb{R}^{+}$ (actually
these functions should be differentiable, so that the
differential formulation of the field theory is not
spoiled). In order to avoid introducing spurious
singularities, we require that $F(t)$ vanish nowhere.
Hence the sign of this function is constant. As a
particular example, the scalar field formulation of
the reduced Gowdy model employed by Pierri
\cite{pierri} (and used afterwards in
Refs.\cite{ccq-t3,torre-prd,come,torre2}) is related
to the CCMV one by a canonical transformation of the
above form with $F(t)=1/\sqrt{t}$ and
$G(t)=-1/(2\sqrt{t})$ \cite{cocome}.

Since the canonical pair $(\varphi,P_{\varphi})$ is
obtained from $(\xi,P_{\xi})$ by a time-dependent
transformation, the classical evolution of these pairs
is different. Completing the Hamiltonian (\ref{hami})
with the time derivative (with respect to the explicit
time dependence) of the generator of the canonical
transformation (\ref{1}) \cite{gold}, one finds that
the evolution of the canonical pair
$(\varphi,P_{\varphi})$ is generated by the ``new
Hamiltonian'' $H_{\varphi}=\oint d\theta {\cal
H}_{\varphi}$ defined by the density
\begin{eqnarray}\label{kami}
{\cal H}_{\varphi}&:=&
\frac{P_{\xi}^2}{2}+\frac{(\xi^{\prime})^2}{2}
+\frac{\dot{F}}{F}\xi P_{\xi}
+\frac{\xi^2}{2}\left(\frac{1}{4t^2}-\dot{G}F+
G\dot{F}\right)\nonumber\\
&=&\frac{P_{\varphi}^2F^2}{2}+
\frac{(\varphi^{\prime})^2}{2F^2}+
\frac{\dot{F}-GF^2}{F}\varphi
P_{\varphi}\nonumber\\
&+&\varphi^2\left\{\frac{G^2}{2}+\frac{1}{2F^2}
\left(\frac{1}{4t^2}-\dot{G}F -G\dot{F}\right)
\right\}.\end{eqnarray} Here, we have not displayed
the time dependence of $F(t)$ and $G(t)$ to simplify
the notation.

As in the case of $(\xi,P_{\xi})$, we fix the
reference time equal to $t_0$ and denote the classical
evolution operator corresponding to the pair
$(\varphi,P_{\varphi})$ by ${\tilde U}(t,t_0)$. In
order to quantize the classical fieldlike variables
$(\varphi,P_{\varphi})$, attaining a unitary
implementation of the corresponding dynamics, we need
to select a complex structure $J_{t_0}$ (on the space
of Cauchy data) at time $t=t_0$ such that
\begin{equation}
\label{u3} {\tilde U}(t,t_0)J_{t_0} {\tilde
U}^{-1}(t,t_0)\sim J_{t_0}, \quad \forall t>0.
\end{equation}
Remembering that, for every symplectic transformation
$A$, $J\sim J'$ if and only if $AJA^{-1}\sim
AJ'A^{-1}$, one can express the unitary
implementability condition (\ref{u3}) in the
equivalent form
\begin{equation}
\label{u7} {\tilde U}(t,t')J_{t_0}{\tilde
U}^{-1}(t,t')\sim J_{t_0}, \quad \forall t, t'>0.
\end{equation}
Similarly, defining $J_t:= {\tilde U}(t,t_0)J_{t_0}
{\tilde U}^{-1}(t,t_0)$, we have
\begin{equation}
\label{u2} J_{t'}={\tilde U}(t',t)J_{t}{\tilde
U}^{-1}(t',t) \sim J_{t},\quad \forall t, t'>0.
\end{equation}

Condition (\ref{u7}), which was the unitary
implementability condition explicitly used in Ref.
\cite{CCMV}, guarantees that the evolution between any
two arbitrarily chosen times is unitary with respect
to $J_{t_0}$ \cite{nota}, whereas condition (\ref{u2})
states that the evolution provides a map between a
family $\{J_t\}$ of equivalent complex structures. In
the present work we follow the standard approach
embodied by Eq. (\ref{u3}), which we take as the
unitary implementation condition.

On the other hand, as we said in Sec. \ref{context},
we will require that the complex structure $J_{t_0}$
be invariant under the gauge group of
$S^1$-translations (\ref{o4}). This is equivalent to
consider only Fock representations for which this
group belongs to the unitary group of the one-particle
Hilbert space, ensuring an invariant unitary
implementation of the gauge group, as in the case
discussed in Sec. \ref{themodel}. We will refer to
such representations as translation invariant
representations, or quantizations.

For the sake of conciseness, in the following we will
discuss only canonical transformations of the type
(\ref{1}) such that the pairs $(\varphi,P_{\varphi})$
and $(\xi,P_{\xi})$ coincide at the fixed reference
time $t_0$. In other words, we will study the case
$F(t_0)=1$ and $G(t_0)=0$. It is not difficult to
realize that this implies no loss of generality. In
fact, any transformation of the form (\ref{1}) can be
decomposed as a time-dependent transformation which
equals the identity at $t=t_0$, and an additional
time-independent transformation with no impact on our
discussion. To be precise, transformation (\ref{1})
can be performed in the following two steps. First, we
introduce the canonical pair:
\begin{equation}
\label{nn2} \tilde \xi:=f(t)\xi, \quad  P_{\tilde
\xi}:=\frac{P_{\xi}}{f(t)}+g(t)\xi,
\end{equation}
with
\begin{equation}
\label{f1g0} f(t_0)=1,\quad g(t_0)=0.
\end{equation}
Secondly, the pair $(\varphi,P_{\varphi})$ is obtained
from $(\tilde\xi,P_{\tilde\xi})$ by a time-independent
transformation \cite{noteZ3}:
\begin{equation}
\label{z2} \varphi=F(t_0) \tilde\xi, \quad
P_{\varphi}=\frac{P_{\tilde\xi}}{F(t_0)}+G(t_0)\tilde\xi.
\end{equation}
It is clear that a quantization of the field theory
described by the pair $(\varphi,P_{\varphi})$ is a
quantization of the system associated with
$(\tilde\xi,P_{\tilde\xi})$, and vice-versa, since the
relation between the two pairs is a local linear
transformation with constant coefficients. In
particular, the coefficients of transformation
(\ref{z2}) are time-independent and
$\theta$-independent. Thus, given a translation
invariant quantization corresponding to one of the
pairs, with unitary dynamics, one immediately obtains
a quantization with the same properties corresponding
to the other pair. The quantum field operators for the
two pairs are of course related by the straightforward
quantum counterpart of Eq. (\ref{z2}), whereas the
quantum evolution operators and translation operators
are actually the same in both cases.

Thus, from now on we will analyze the consequences of
demanding a unitary implementation of the dynamics for
the pair $(\tilde\xi,P_{\tilde\xi})$, with respect to
translation invariant Fock representations. After the
derivation of the unitary implementability condition
in explicit form, the proof of our uniqueness result
will be split into two parts. We will first show that
a unitary dynamics for the pair
$(\tilde\xi,P_{\tilde\xi})$ can be achieved only if
the function $f(t)$ in Eq. (\ref{nn2}) is the constant
unit function [equivalently, unitary dynamics for
$(\varphi,P_{\varphi})$ is reached only if the
function $F(t)$ in Eq. (\ref{1}) is constant]. We will
then prove the uniqueness of the quantization for
those cases in which unitarity is attained.

Note that the CCMV quantization already provides a
representation of the time $t_0$-fields corresponding
to the pairs $(\tilde\xi,P_{\tilde\xi})$ and
$(\varphi,P_{\varphi})$. Clearly, in the case of
$(\tilde\xi,P_{\tilde\xi})$ the $t_0$-quantum fields
coincide with the CCMV ones, whereas in the
$(\varphi,P_{\varphi})$ case the (Fourier components
of the) fields are related by $\hat{\varphi}_n(t_0)=
F(t_0)\hat{\xi}_n(t_0)$ and
$\hat{P}_{\varphi}^n(t_0)=[1/F(t_0)]\hat{P}_{\xi}^n(t_0)
+G(t_0)\hat{\xi}_n(t_0)$, where $\hat{\xi}_n(t_0)$ and
$\hat{P}_{\xi}^n(t_0)$ are the CCMV operators. We will
see that the dynamics of the pair
$(\tilde\xi,P_{\tilde\xi})$ with $f(t)=1$ [or
$(\varphi,P_{\varphi})$ with constant $F(t)$] is
unitarily implementable in the CCMV representation.
Most importantly, we will show that whenever the
dynamics of $(\tilde\xi,P_{\tilde\xi})$ can be
implemented unitarily, the corresponding translation
invariant Fock representation also provides a unitary
implementation of the dynamics of the pair
$(\xi,P_{\xi})$, and is therefore unitarily equivalent
to the CCMV representation by the results of Ref.
\cite{CCMV}.

\section{Unitarity condition}
\label{compleuni}

Let us consider then the field description
corresponding to the canonical pair $(\tilde
\xi,P_{\tilde\xi})$ (\ref{nn2}), with the real
functions $f(t)$ and $g(t)$ satisfying conditions
(\ref{f1g0}). Note that, given the continuity and
non-vanishing of $f(t)$, we now have $f(t)>0$ $\forall
t>0$.

As in the case of $(\xi,P_{\xi})$, we perform the
Fourier decomposition (\ref{o5}) for the new canonical
pair $(\tilde \xi,P_{\tilde \xi})$, and introduce
corresponding variables $\{{\tilde B}_m(t)\}$, like in
Eqs. (\ref{o6},\ref{obcal}). In agreement with our
above remarks, we note that the set $\{{\tilde
B}_m\}:=\{{\tilde B}_m(t_0)\}$ coincides with
$\{B_m\}$ since our canonical transformation is the
identity at the reference time. Thus, the same
kinematical variables $\{B_m\}$ are used in the
quantization of the two field descriptions of the
model, $(\xi,P_{\xi})$ and $(\tilde \xi,P_{\tilde
\xi})$. The classical evolution in the new description
is different, as we have commented. From the
definition of $\{{\tilde B}_m(t)\}$ and Eqs.
(\ref{nn2},\ref{f1g0}), one can check that the
evolution matrices ${\cal U}_m(t,t_0)$ introduced in
Eq. (\ref{o15}) are now replaced with
\begin{equation} \label{3} \tilde {\cal U}_m(t,t_0)
= C_m(t){\cal U}_m(t,t_0),
\end{equation}
where
\begin{eqnarray} \label{2} C_m(t) &:=& \frac{1}{2}
\left(\begin{array}{cc} f_{+}(t)+i\frac{g(t)}{m} \;&
f_{-}(t)+i\frac{g(t)}{m} \\ f_{-}(t)-
i\frac{g(t)}{m}\;& f_{+}(t)-i\frac{g(t)}{m}
\end{array} \right),\\
f_{\pm}(t)&:=&f(t)\pm \frac{1}{f(t)}.
\end{eqnarray}
The matrices $C_m(t)$ actually describe the canonical
transformation (\ref{nn2},\ref{f1g0}) in the variables
$\{B_m(t)\}$, and so $C_m(t_0)={\bf 1}$.

A straightforward calculation shows that
\begin{equation}
\label{tildeu}
\tilde {\cal U}_m(t,t_0)= \left(
\begin{array}{cc} \tilde \alpha_m(t,t_0) &
\tilde \beta_m(t,t_0) \\ \tilde \beta_m^*(t,t_0) &
\tilde \alpha_m^*(t,t_0)
\end{array} \right),
\end{equation}
with
\begin{eqnarray}
\label{4} 2 \tilde \alpha_m(t,t_0)  & := &
f_+(t)\alpha_m(t,t_0) + f_-(t)\beta^*_m(t,t_0)
\nonumber \\
& + & i
\frac{g(t)}{m}[\alpha_m(t,t_0)+\beta_m^*(t,t_0)],\\
\label{5} 2 \tilde \beta_m(t,t_0)  & := & f_+(t)
\beta_m(t,t_0) + f_-(t)\alpha^*_m(t,t_0)
\nonumber \\
& + &
i\frac{g(t)}{m}[\alpha^*_m(t,t_0)+\beta_m(t,t_0)],
\end{eqnarray}
where $\alpha_m$ and $\beta_m$ are defined in Eq.
(\ref{o15}).

According to our previous comments, in order to
achieve an admissible quantization of the fieldlike
variables $(\tilde \xi,P_{\tilde \xi})$, one looks for
complex structures $J$ (at time $t_0$) that are
invariant under $S^1$-translations and lead to a
unitary implementation of the evolution given by Eq.
(\ref{tildeu}) $\forall t>0$. At this point, one can
employ a result proven in Ref. \cite{CCMV}, namely,
that every such invariant complex structure $J$ is
related to $J_0$ by a symplectic transformation, where
$J_0$ is the complex structure used in the CCMV
quantization of Refs. \cite{cocome2,cocome}.
Explicitly, every invariant complex structure can be
expressed as $J=K_JJ_0K_J^{-1}$, where $K_J$ is block
diagonal in the basis $\{B_m\}$, with $4\times 4$
blocks of the form \begin{eqnarray} (K_J)_m&=&\!\left(
\begin{array}{cc}
({\cal K}_J)_m & {\bf 0}  \\
{\bf 0} & ({\cal K}_J)_m
\end{array} \right),\; ({\cal K}_J)_m= \left(
\begin{array}{cc}\kappa_m & \lambda_m  \\
\lambda_m^* & \kappa_m^*\end{array} \right),
\nonumber\\ \label{nn1}
|\kappa_m|^2&=&1+|\lambda_m|^2.
\end{eqnarray}

On the other hand, a symplectic transformation $A$
admits a unitary implementation with respect to a
complex structure $J=K_JJ_0K_J^{-1}$ if and only if
$K_J^{-1}AK_J$ is unitarily implementable with respect
to $J_0$. Thus, the condition for unitary
implementation of the classical dynamics
(\ref{tildeu}) is that the antilinear part of the
symplectic transformation defined by the matrices
\[({\cal K}_J)_m^{-1}\tilde {\cal U}_m(t,t_0)({\cal
K}_J)_m= ({\cal K}_J)_m^{-1}C_m(t){\cal
U}_m(t,t_0)({\cal K}_J)_m \] be Hilbert-Schmidt in the
Hilbert space defined by $J_0$, $\forall t>0$. This in
turn translates into the following square summability
condition: the dynamics of the fieldlike variables
$(\tilde \xi,P_{\tilde \xi})$ has a unitary
implementation with respect to an invariant complex
structure $J=K_JJ_0K_J^{-1}$ if and only if the
sequence $\{\tilde\beta^J_m(t,t_0)\}$, with
\begin{eqnarray}
\label{6}\tilde\beta^J_m(t,t_0)&:=&
(\kappa_m^*)^2\tilde\beta_m(t,t_0)
-\lambda_m^2\tilde\beta_m^*(t,t_0)
\nonumber\\
&+&2i \kappa_m^*\lambda_m{\rm
Im}[\tilde\alpha_m(t,t_0)],
\end{eqnarray}
(where ${\rm Im}$ denotes the imaginary part) is
square summable for all strictly positive $t$, i.e. if
and only if the sum
$\sum_{m=1}^{\infty}|\tilde\beta^J_m(t,t_0)|^2$ exists
$\forall t>0$.

\section{``No-go'' result for time-dependent scalings}
\label{nogo}

We will now prove that, for the sequence
$\{\tilde\beta^J_m(t,t_0)\}$ to be square summable, it
is necessary that the scaling function $f(t)$ in the
transformation (\ref{nn2}) be constant. As we will
see, the square summability condition fails strongly
otherwise, in the sense that $\tilde\beta^J_m(t,t_0)$
does not even go to zero $\forall t>0$ when $m\to
\infty$. We will therefore obtain $f(t)=f(t_0)=1$
$\forall t>0$ as a necessary condition for unitarity.

Let us consider the related sequence
$\{\tilde\beta^J_m(t,t_0)/(\kappa_m^*)^2\}$. Since
$|\kappa_m|^2\geq 1$ by Eq. (\ref{nn1}), we have
\begin{equation}
\left|\frac{\tilde\beta^J_m(t,t_0)}{(\kappa_m^*)^2}
\right|^2\leq |\tilde\beta^J_m(t,t_0)|^2,\quad \forall
m\in \mathbb{N},\;\forall t>0.\end{equation} Therefore
$\{\tilde\beta^J_m(t,t_0)/(\kappa_m^*)^2\}$ must be
square summable whenever $\{\tilde\beta^J_m(t,t_0)\}$
is. In particular, a necessary condition for unitarity
is that $\tilde\beta^J_m(t,t_0)/(\kappa_m^*)^2$ tend
to zero in the limit $m\to\infty$, $\forall t>0$. We
will now analyze the consequences of this condition.

Using again Eq. (\ref{nn1}), we conclude that
$|\lambda_m /\kappa_m|\leq 1$ $\forall m\in
\mathbb{N}$. Then, one can check from Eq. (\ref{6})
that all the time-independent factors appearing in
$\tilde\beta^J_m(t,t_0)/(\kappa_m^*)^2$ are bounded.
On the other hand, since $|\alpha_m(t,t_0)|$ and
$|\beta_m(t,t_0)|$ have well defined limits when $m\to
\infty$ for every fixed $t$ [see discussion below Eq.
(\ref{o15})], they also form bounded sequences for
each $t>0$. As a consequence, the contribution of the
terms that contain $g(t)$ in Eqs. (\ref{4},\ref{5})
[which provide $\tilde{\alpha}_m(t,t_0)$ and
$\tilde{\beta}_m(t,t_0)$] are (at most) of order
$1/m$. Hence, the corresponding contribution in $g(t)$
to $\tilde\beta^J_m(t,t_0)/(\kappa_m^*)^2$ is also of
this order and thus tends to zero when $m\to\infty$.
This means that, up to corrections of order $1/m$ that
are negligible for large $m$, we can work with the
approximation
\begin{eqnarray}
2\tilde{\alpha}_m(t,t_0)& \approx &
f_+(t)\alpha_m(t,t_0) +
f_-(t)\beta^*_m(t,t_0),\nonumber\\ \label{g05}
2\tilde{\beta}_m(t,t_0)& \approx &
f_+(t)\beta_m(t,t_0) + f_-(t)\alpha^*_m(t,t_0).
\end{eqnarray}
Let us consider the dominant terms of these
expressions when $m\to \infty$, and let us call them
$\tilde{\alpha}^0_m$ and $\tilde{\beta}^0_m$. They can
be easily deduced using that $\alpha_m-e^{-imT}$ and
$\beta_m$ tend to zero in this limit, where $T:=t-t_0
>-t_0$. Thus,
\begin{eqnarray}
\label{11} 2\tilde{\alpha}^0_m(t,t_0)&=&
f_+(t)e^{-imT},\nonumber\\
2\tilde{\beta}^0_m(t,t_0)&=& f_-(t)e^{imT}.
\end{eqnarray}
Employing again that the time-independent coefficients
entering $\tilde\beta^J_m(t,t_0)/(\kappa_m^*)^2$ are
bounded, we conclude that this sequence vanishes in
the limit $m\to \infty$ if and only if so does the
corresponding sequence obtained by replacing $\tilde
\alpha_m$ and $\tilde \beta_m$ with
$\tilde{\alpha}^0_m$ and $\tilde{\beta}^0_m$, namely
the sequence with elements
\begin{eqnarray}
\label{zero9}
\frac{\tilde{\beta}^{0J}_m(t,t_0)}{(\kappa_m^*)^2}&:=&
\tilde{\beta}_m^0(t,t_0)-\frac{\lambda_m^2}
{(\kappa_m^*)^2}\tilde{\beta}_m^{0*}(t,t_0)
\nonumber\\
&+&2i{\frac{\lambda_m}{\kappa_m^*}}{\rm
Im}[\tilde{\alpha}_m^0(t,t_0)].
\end{eqnarray}
Hence, as a necessary condition for a unitary
implementation of the dynamics,
$\tilde{\beta}^{0J}_m(t,t_0)/(\kappa_m^*)^2$ must tend
to zero when $m\to\infty$.

By substituting expressions (\ref{11}) for
$\tilde{\alpha}_m^0$ and $\tilde{\beta}_m^{0}$, we
then conclude that the following two real sequences
(with $m\in \mathbb{N}$), which give the real and
imaginary parts of
$\tilde{\beta}^{0J}_m(t,t_0)/(\kappa_m^*)^2$, must
vanish in the limit $m\to\infty$ $\forall t>0$:
\begin{eqnarray}
\label{n7}&& \left({\rm
Im}\left[\frac{\lambda_m}{\kappa^*_m}\right] f_+(t) -
{\rm Im}\left[\frac{\lambda_m^2}
{(\kappa^*_m)^2}\right]
\frac{f_-(t)}{2}\right)\sin(mT)\nonumber\\
&&+\left(1-{\rm
Re}\left[\frac{\lambda_m^2}{(\kappa^*_m)^2}
\right]\right)\frac{f_-(t)}{2}\cos(mT) ,
\end{eqnarray}
and
\begin{eqnarray}
\label{n8} &&\left(\left\{1+ {\rm Re}
\left[\frac{\lambda_m^2}{(\kappa^*_m)^2}\right]
\right\}\frac{f_-(t)}{2}- {\rm
Re}\left[\frac{\lambda_m}
{\kappa^*_m}\right]f_+(t)\right) \nonumber\\
&&\times\sin(mT)-{\rm Im}
\left[\frac{\lambda_m^2}{(\kappa^*_m)^2}\right]
\frac{f_-(t)}{2}\cos(mT).
\end{eqnarray}
Here, the symbol ${\rm Re}$ denotes the real part.

Actually, as we show in Appendix \ref{proof}, it is
impossible that the imaginary part of
$\tilde{\beta}^{0J}_m(t,t_0)/(\kappa^*_m)^2$ [given by
Eq. (\ref{n8})] tends to zero $\forall t>0$, as
required, if the time-independent coefficients of the
cosine terms in the above expressions,
\begin{equation}\label{coscoef} 1-{\rm
Re}\left[\frac{\lambda_m^2}{(\kappa^*_m)^2}\right]\quad
{\rm and}\quad {\rm Im}\left[\frac{\lambda_m^2}{
(\kappa^*_m)^2}\right],\end{equation} tend to zero
simultaneously on any subsequence $S\subset
\mathbb{N}$ (i.e. for $m\in S\subset \mathbb{N}$).
This places us in an adequate position to prove that a
necessary condition for the dynamics of the fieldlike
variables $(\tilde \xi, P_{\tilde \xi})$ to admit a
unitary implementation with respect to some invariant
complex structure is that the scaling function $f$ be
constant.

Let us start by taking $T=2\pi q/p$, where $q$ and $p$
are arbitrary integers subject only to the condition
that $2\pi q/p>-t_0$. For each fixed $p$, we then
consider the subsequence $S_p:=\{m=np,\ n\in
\mathbb{N}\}$. Since the terms (\ref{n7}) and
(\ref{n8}) tend to zero when $m\to \infty$ $\forall
T>-t_0$ ($t>0$), the same happens on each $S_p$ for
every $q$. Thus, taking into account that $\sin(2\pi n
q)=0$ and $\cos(2\pi n q)=1$, one obtains that both
\begin{equation}
\left(1-{\rm Re}
\left[\frac{\lambda_{np}^2}{(\kappa^*_{np})^2}\right]
\right)f_-\left(t_0+\frac{2\pi q}{p}\right)
\end{equation}
and
\begin{equation}
 {\rm Im}
\left[\frac{\lambda_{np}^2}{(\kappa^*_{np})^2}\right]
f_-\left(t_0+\frac{2\pi q}{p}\right)
\end{equation}
must tend to zero as $n\to\infty$ for all possible
values of $p$ and $q$. However, since we know that the
time-independent coefficients in these expressions
cannot have simultaneously a zero limit on any
subsequence $S_p$ [see Appendix \ref{proof}], our
conditions can only be fulfilled if $f_{-}(t_0+2\pi
q/p)$ vanishes $\forall p,q$ or, equivalently, if
\begin{equation}
\label{19} f^2\left(t_0+\frac{2\pi q}{p}\right)=1,
\quad \forall q, p.
\end{equation}
But, given that the set $\{t_0+2\pi q/p\}$ is dense on
the half-line of positive numbers and $f^2(t)$ is a
continuous function, this implies that $f^2(t)$ must
be the unit constant function. Using again the
continuity of $f(t)$ and that $f(t_0)=1$, we then see
that $f(t)$ itself must be the unit function. This
ends our proof.

In conclusion, we have shown that, with a
(translation) invariant complex structure, no unitary
implementation of the dynamics can be achieved unless
transformation (\ref{nn2}) is actually a simple
redefinition of the momentum:
\begin{equation}
\label{nn2f1} \tilde \xi= \xi, \quad  P_{\tilde \xi}=
P_{\xi}+g(t)\xi.\end{equation}

Let us end the section with the following remark. If
we now turn to the general field parameterization
(\ref{1}), it follows  from our comments in Sec.
\ref{alternate} that a necessary condition for a
unitary implementation of the corresponding dynamics
is that the function $F(t)$ in Eq. (\ref{1}) be
constant, $F(t)=F(t_0)$ $\forall t>0$ \cite{noteZ4}.
Thus, one can already conclude that the CCMV choice of
fundamental field $\xi$ for the reduced Gowdy model
(and ignoring for the moment the choice of momentum)
is essentially unique if a unitary dynamics is to be
achieved. No time-dependent scaling of this field is
allowed. In particular, this shows that the field
version employed by Pierri \cite{pierri} admits no
unitary implementation of the dynamics with respect to
any of all the possible invariant complex structures.

\section{Equivalence of representations}
\label{equivrepre}

We will now focus our discussion on the remaining
transformations (\ref{nn2f1}) and show that the
dynamics of the fieldlike variables $(\tilde \xi,
P_{\tilde \xi})$ is unitary if and only if so is the
dynamics of $(\xi, P_{\xi})$.

For this unitarity, it is still necessary that
expressions (\ref{n7},\ref{n8}), now particularized to
$f(t)=1$ [i.e. $f_+(t)=2$ and $f_-(t)=0$], tend to
zero when $m\to \infty$ for all strictly positive
values of $t$. We then arrive at the necessary
conditions
\begin{eqnarray}
\label{nn5} {\rm Im}\left[\frac{\lambda_m}{
\kappa^*_m}\right]\sin(mT)&\rightarrow& 0,\\
\label{nn6}
 {\rm Re}\left[\frac{\lambda_m}{
\kappa^*_m}\right]\sin(mT)&\rightarrow& 0
\end{eqnarray}
for every $T>-t_0$. Thus, in order to avoid the false
conclusion that $\sin^2(mT)$ goes to zero on a
subsequence of positive integers e.g. $\forall T\in
[0,2\pi]$ (like in the calculations explained in
Appendix \ref{proof}), it is necessary that both ${\rm
Im}[\lambda_m/ \kappa^*_m]$ and ${\rm Re}[\lambda_m/
\kappa^*_m]$ tend to zero. So,
${|\lambda_m|^2/|\kappa_m|^2}$ must vanish in the
limit $m\to\infty$. Using Eq. (\ref{nn1}), this means
that $1/|\kappa_m|^2$ must approach the unit, what
implies that the sequence $\{\kappa_m\}$ has to be
bounded.

Let us then start again from Eq. (\ref{6}), analyzing
the original condition of square summability of
$\{\tilde\beta^J_m\}$ required for the unitary
implementation of the dynamics. Recalling again Eq.
(\ref{nn1}) and the fact that the sequence
$\{\kappa_m\}$ is bounded, we see that all
time-independent coefficients appearing in expression
(\ref{6}) for $\tilde\beta^J_m$ are bounded. In
addition, from Eqs. (\ref{4},\ref{5}) with $f(t)= 1$,
we have
\begin{eqnarray}
\tilde{\alpha}_m(t,t_0) &=& \alpha_m(t,t_0) +
i\frac{g(t)}{2m}
[\alpha_m(t,t_0)+\beta_m^*(t,t_0)],\nonumber \\
\tilde{\beta}_m(t,t_0)&=&\beta_m(t,t_0) +
i\frac{g(t)}{2m} [\alpha^*_m(t,t_0)+\beta_m(t,t_0)].
\nonumber\end{eqnarray} When the above expressions are
introduced in Eq. (\ref{6}) for $\tilde{\beta}^J_m$,
one immediately sees that the contribution of terms in
$g(t)$ are automatically square summable, owing to the
fact that all terms proportional to $g(t)$ come with a
factor of $1/m$, that the time-independent
coefficients in $\tilde{\beta}^J_m$ are bounded, and
that $\alpha_m(t,t_0)$ and $\beta_m(t,t_0)$ are also
bounded $\forall t>0$. The condition for a unitary
dynamics is then the square summability of the
remaining contribution to $\tilde\beta^J_m$, namely
\begin{eqnarray}
\label{xi6}
\beta^J_m(t,t_0)&:=&(\kappa_m^*)^2\beta_m(t,t_0)
-\lambda_m^2\beta_m^*(t,t_0)\nonumber\\
&+&2i \kappa_m^*\lambda_m{\rm Im}[\alpha_m(t,t_0)].
\end{eqnarray}
But this term $\beta^J_m(t,t_0)$ is precisely the
$\beta$-coefficient corresponding to the antilinear
part of the classical evolution operator (\ref{o15})
for the canonical pair $(\xi,P_{\xi})$ with the choice
of complex structure $J=K_J J_0 K_J^{-1}$ \cite{CCMV}.
Therefore, the dynamics of the pair $(\tilde
\xi,P_{\tilde \xi})$ is unitarily implementable with
respect to an invariant complex structure if and only
if the dynamics of the CCMV fieldlike variables
$(\xi,P_{\xi})$ admits a unitary implementation with
respect to the same structure. One can now invoke the
results of Ref. \cite{CCMV}, where it was proven that
any invariant complex structure $J$ which allows a
unitary implementation of the dynamics of
$(\xi,P_{\xi})$ provides a quantum representation
which is unitarily equivalent to that determined by
$J_0$, i.e. the CCMV representation constructed in
Refs. \cite{cocome,cocome2}.

Summarizing, we have demonstrated that there is a
unique (equivalence class of) translation invariant
Fock representation(s) of the fields at time $t=t_0$
such that the evolution of the canonical pair of
fields given by transformation (\ref{nn2f1}), for any
function $g(t)$, is unitary implementable. This
representation is the one constructed in Refs.
\cite{cocome,cocome2} and is determined by the complex
structure $J_0$. Furthermore, as explained in Sec.
\ref{alternate}, this conclusion applies as well to
any field parameterization defined by a transformation
of the form (\ref{1}) with a nonnegative constant
function $F(t)=F(t_0)$ and any function $G(t)$.

In particular, no new quantum representations appear
when one looks for unitary implementations of the
dynamics of the transformed canonical pair $(\tilde
\xi,P_{\tilde \xi})$. The quantization defined by
$J_0$ already gives a unitary implementation of such
dynamics, and there are no more (inequivalent)
quantizations.

It is worth noticing that, on general grounds, given
any representation which allows a unitary dynamics for
the two canonical pairs $(\xi,P_{\xi})$ and
$(\tilde\xi,P_{\tilde\xi})$, there is a well defined
quantum version of the momentum redefinition
(\ref{nn2f1}) provided by the time-dependent unitary
operator $\hat{U}^{-1}(t,t_0)\hat{\tilde U}(t,t_0)$,
where $\hat U(t,t_0)$ and $\hat{{\tilde U}}(t,t_0)$
are, respectively, the quantum evolution operators
corresponding to the dynamics of the pairs
$(\xi,P_{\xi})$ and $(\tilde\xi,P_{\tilde\xi})$. Our
result is, however, much stronger: different field
descriptions are not only unitarily related for a
given representation, but there is actually a unique
(equivalence class of) translation invariant
representation(s) admitting a unitary dynamics.

\section{Summary and conclusions}
\label{summary}

We have analyzed the uniqueness of the Fock
quantization of the family of linearly polarized Gowdy
$T^3$ cosmologies after its reduction by a
gauge-fixing procedure which removes all the
constraints except for a homogeneous one. This
constraint generates translations on the coordinate
$\theta\in S^1$ that, together with the time
coordinate $t$, parameterize the set of orbits of the
isometry group. The phase space of this reduced model
can be viewed as that corresponding to a
point-particle degree of freedom and a scalar field.
With a suitable parameterization of the induced
metric, this field satisfies a Klein-Gordon equation
on a fiducial flat 1+1 background subject to a
time-dependent potential, which is invariant under the
gauge group of $S^1$-translations. Besides, one can
choose the canonical momentum of this field in such a
way that the Hamiltonian density that generates the
dynamics is quadratic both in the field and in its
momentum (without crossed terms): this is the CCMV
field formulation introduced in Refs.
\cite{cocome2,cocome} for the description of the phase
space of the reduced Gowdy model.

In a previous work, it was shown that the Fock
quantization of this field formulation, which depends
on the choice of complex structure, is unique under
some natural requirements. More precisely, if one
demands that the complex structure be invariant under
$S^1$-translations, so that every element of the gauge
group is represented by a unitary operator that leaves
the Fock vacuum invariant, then any Fock quantization
admitting a unitary implementation of the field
dynamics is unitarily equivalent to the CCMV
quantization, which was obtained with a particular
choice of complex structure $J_0$. In the present
paper we have extended this uniqueness result to cover
all reasonable Fock quantizations of the reduced Gowdy
model by considering also the freedom available in the
choice of the field description of the system.
Specifically, we have studied local field
reparameterizations of the induced metric in the
reduced model which are independent of the spatial
coordinates (so that they commute with the isometry
and gauge groups), respect the decoupling with the
point-particle degrees of freedom (attained in the
CCMV parameterization), and whose dynamics is governed
by a homogeneous Klein-Gordon type field equation.
Such reparameterizations amount to a time-dependent
scaling of the scalar field. Its canonical momentum is
scaled by the inverse factor and, in principle, may
also get a time-dependent linear contribution in the
field.

We have concentrated our discussion in the case when
such a linear, time-dependent canonical transformation
of the CCMV variables coincides with the identity at
the fixed reference time $t=t_0$, which determines the
Cauchy surface with respect to which the quantum
representation of the fields is constructed. The most
general situation can be obtained from this case by
combining it with a time-independent canonical
transformation which produces constant linear
combinations of the $t_0$-fields and does not affect
the conclusions about uniqueness. For the case of
time-depend\-ent transformations which are the
identity at $t_0$, we have then proven that the new
canonical pair of fieldlike variables admits a Fock
quantization, defined by an invariant complex
structure (under $S^1$-translations) and providing a
unitary implementation of the field dynamics, if and
only if the scaling function is the unit function. In
particular, this demonstrates once and for all that
there exists no Fock quantization with these
properties for the scalar field formulation of the
reduced model adopted by Pierri \cite{pierri,come}.

Moreover, even in the remaining case of no scaling
(i.e. a unit scaling function), where only the
canonical momentum differs from that of the CCMV
description [see Eq. (\ref{nn2f1})], we have shown
that the Fock representation of the $t_0$-fields
corresponding to the transformed canonical pair is
unique, in the sense that, if it is defined by an
invariant complex structure and admits a unitary
dynamics, it is unitarily equivalent to the CCMV
representation determined by the complex structure
$J_0$. No new (inequivalent) translation invariant
representations with unitary dynamics appear by
adopting a canonical momentum different from that of
the CCMV formulation.

Furthermore, it is possible to eliminate the freedom
in the choice of canonical momentum by including an
additional requirement on the quantization. Namely,
one further demands that there exists a choice of
complex structure such that the Fock vacuum of the
corresponding representation belongs to the domain of
the generator of the evolution. This condition is
convenient in practice, because it allows one to
calculate the action of the evolution operator on the
vacuum (and on the $n$-{\it particle} states) by
expanding it in powers of the generator. Appendix
\ref{vacham} shows that, with this additional demand,
one can actually fix the canonical pair of fieldlike
variables so that it coincides with the CCMV pair.

Therefore, we conclude that all Fock quantizations
obtained with a reasonable field description of the
reduced Gowdy model are unitarily equivalent under
natural requirements. In this sense, the CCMV
quantization of the Fock type introduced in Refs.
\cite{cocome2,cocome} is unique. On the other hand, if
one considered instead the unreduced Gowdy model [see
metric (\ref{metricunred})], rather than its
gauge-fixed and reduced version, there would still be
freedom in the choice of gauge. Nonetheless, the gauge
adopted is certainly well motivated both from a
geometrical and a physical point of view. The
$\theta$-diffeomorphism gauge freedom has been fixed,
except for the group of translations, by requiring the
homogeneity of the phase space variable that generates
conformal transformations of the two-metric induced on
the set of group orbits. The homogeneous part of this
variable is a known Dirac observable of the Gowdy
cosmologies (i.e. it commutes with all the constraints
of the unreduced model) \cite{mizo}. In addition, the
phase-space variable chosen as time coordinate, apart
from a multiplicative factor, is the area of the
orbits of the group of isometries, which expands
monotonously in the evolution of the cosmological
solutions and whose gradient has a timelike character
that is invariant under coordinate transformations.

Finally, it is worth emphasizing that our uniqueness
result provides an example of a cosmological system in
which, without abandoning standard quantum field
theory, one can single out a preferred quantization by
requiring suitable symmetry and consistency
conditions. In the considered case of the reduced
Gowdy model, this strongly supports the conclusion
that the physical consequences that can be derived
from the CCMV quantization are meaningful and not an
artifact of the scalar field description and Fock
representation adopted for the system.

\section*{Acknowledgements}

The authors are thankful to A. Ashtekar, M.
Varadarajan, L.J. Garay, and M. Mart\'{\i}n-Benito for
helpful conversations and discussions. This work was
supported by the Spanish MEC Projects
FIS2005-05736-C03-02 and FIS2006-26387-E/, and the
Portuguese FCT Project POCTI/FIS/57547/2004.

\appendix

\section{A proof for time-dependent scalings}
\setcounter{equation}{0}
\renewcommand{\theequation}{A\arabic{equation}}
\label{proof}

We want to prove that, if
\begin{equation}\label{coscoef2} 1-{\rm
Re}\left[\frac{\lambda_m^2}{(\kappa^*_m)^2}\right]\quad
{\rm and}\quad {\rm Im}\left[\frac{\lambda_m^2}{
(\kappa^*_m)^2}\right]\end{equation} tend both to zero
on a subsequence $S\subset \mathbb{N}$ (i.e. for $m\in
S\subset \mathbb{N}$), it is impossible that the
imaginary part of
$\tilde{\beta}^{0J}_m(t,t_0)/(\kappa^*_m)^2$ has a
vanishing limit $\forall t>0$. We recall that
expression (\ref{coscoef2}) provides the
time-independent coefficients of the cosine terms of
the real and imaginary parts of
$\tilde{\beta}^{0J}_m(t,t_0)/(\kappa^*_m)^2$, given by
Eqs. (\ref{n7},\ref{n8}). Let us remind also that
$t_0>0$ is fixed and that we call the difference of
times $T:=t-t_0$.

In order to prove our statement, we first note that,
if the two coefficients (\ref{coscoef2}) tend to zero
on certain subsequence $S\subset \mathbb{N}$, then
$\left({\rm Re}[\lambda_m/ \kappa^*_m]\right)^2$ must
tend to $1$ on $S$. This can be seen by summing the
square of the two coefficients for each $m\in S$,
which gives $(1-|\lambda_m/
\kappa^*_m|^2)^2+4\left({\rm Im}[\lambda_m/
\kappa^*_m]\right)^2$. Since this expression must tend
to zero on $S$, we get that $|\lambda_m/ \kappa^*_m|$
tends to 1 and ${\rm Im}[\lambda_m/ \kappa^*_m]$ to
zero. But then $\left({\rm Re}[\lambda_m/
\kappa^*_m]\right)^2$ tends to $1$ on $S$ as we
anticipated.

Let us then suppose that when $m\to \infty$, the
imaginary part of
$\tilde{\beta}^{0J}_m(t,t_0)/(\kappa^*_m)^2$,
displayed in Eq. (\ref{n8}), vanishes $\forall t>0$.
Thus, it does so on any possible subsequence of
positive integers $m$. Let us now suppose that there
is a particular subsequence $S\subset \mathbb{N}$ such
that the coefficients (\ref{coscoef2}) tend both to
zero on $S$. Given that the term which multiplies
${\rm Im}[\lambda_m^2/(\kappa^*_m)^2]$ in Eq.
(\ref{n8}), namely $f_-(t) \cos(mT)/2$, is bounded for
every particular value of $t$, we conclude that
\[\left(\left\{1+ {\rm Re}
\left[\frac{\lambda_m^2}{(\kappa^*_m)^2}\right]
\right\}\frac{f_-(t)}{2}- {\rm
Re}\left[\frac{\lambda_m}
{\kappa^*_m}\right]f_+(t)\right)\sin(mT)\] must have a
zero limit on $S$, $\forall t>0$. Moreover, since
$1-{\rm Re}\left[\lambda_m^2/(\kappa^*_m)^2\right]$
also tends to zero on $S$, we get that
\begin{equation}
\label{n13} \left( -{\rm Re}\left[\frac{\lambda_m}{
\kappa^*_m}\right]f_+(t) +f_-(t)\right)\sin(mT)
\end{equation}
must tend to zero on $S$ $\forall t>0$.

In addition, as we have seen above, $\left({\rm
Re}[\lambda_m/ \kappa^*_m]\right)^2$ necessarily tends
to 1 on $S$. Then, there exists at least one
subsequence $S'\subset S$ such that ${\rm
Re}[\lambda_m/ \kappa^*_m]$ tends to $1$ or to $-1$ on
$S'$. In any of these cases, given that $S'\subset S$,
the sequence (\ref{n13}) must tend to zero on $S'$ and
(recalling the definition of $f_{\pm}$) we obtain that
either $\sin(mT)f(t)$ or $\sin(mT)/f(t)$ (or both)
have a zero limit on some subsequence $S'\subset
\mathbb{N}$ $\forall t>0$. Thus, since $f(t)$ is
continuous and vanishes nowhere, $\sin(mT)$ must tend
to zero on $S'$ $\forall t>0$, and therefore $\forall
T>-t_0$. In particular, this implies that $\sin^2(mT)$
tends to zero on $S'$ $\forall T\in[0,2\pi]$. However,
this last conclusion is false. For instance, Lebesgue
dominated convergence \cite{kolfo} would then imply
that $\int_0^{2\pi}dT \sin^2(mT)$, which is clearly
equal to $\pi$ for all nonzero integers $m$, has to
converge to zero on $S'$. This indicates a
contradiction. Therefore, since the imaginary part of
$\tilde{\beta}^{0J}_m(t,t_0)/(\kappa^*_m)^2$ must tend
to zero, one can exclude the possibility that the two
sequences of time-independent coefficients appearing
in Eq. (\ref{coscoef2}) can both converge to zero on
any subsequence $S\subset \mathbb{N}$.

\section{A criterion for the choice of canonical
momentum} \setcounter{equation}{0}
\renewcommand{\theequation}{B\arabic{equation}}
\label{vacham}

We have seen  that there exists some freedom in the
definition of the momentum canonically conjugate to
the CCMV field $\xi$ [see Eq. (\ref{nn2f1})], although
this freedom does not result in the availability of
new (inequivalent) Fock quantizations for the reduced
Gowdy model. We will now introduce a possible
criterion to remove this freedom and select a
preferred canonical momentum.

Our starting point is a time-dependent canonical
transformation of the form $\tilde{\xi}=\xi$ and
$P_{\tilde{\xi}}=P_{\xi}+g(t)\xi$ where the function
$g(t)$ is (at least) continuous and vanishes at the
reference time $t_0$. In addition to our requirements
of invariance under $S^1$-translations and unitarity
of the dynamics, we will demand that the complex
structure $J$ that determines the Fock representation
for the canonical pair $(\tilde{\xi},P_{\tilde{\xi}})$
be such that the associated vacuum belongs to the
domain of the generator of the evolution in the
Schr\"odinger picture. This additional requirement on
the vacuum is of practical interest since it is
necessary to render meaningful the action of the
evolution operator (in the Schr\"odinger picture) on
the dense subspace formed by the $n$-{\it particle}
states when one expands this operator as a formal
series in powers of its generator.

The classical generator $H_{\tilde{\xi}}$ of the
dynamics of the canonical pair
$(\tilde{\xi},P_{\tilde{\xi}})$ can be easily obtained
from Eq. (\ref{kami}) by setting $F(t)=1$ and
$G(t)=g(t)$. In terms of the CCMV pair, this generator
reads
\begin{equation}\label{gene}
H_{\tilde{\xi}}=\frac{1}{2}\oint d\theta \left[
P_{\xi}^2+(\xi^{\prime})^2+\xi^2 \left(\frac{1}{4
t^2}-\dot{g}(t)\right)\right].
\end{equation}
In the basis $\{B_m(t)\}$  introduced in Eq.
(\ref{obcal}), the classical generator is thus
$H_{\tilde{\xi}}=H_{\tilde{\xi}}^0+
H_{\tilde{\xi}}[t|\{B_m(t)\}]$ where
$H_{\tilde{\xi}}^0$ denotes the contribution of the
zero modes and \cite{CCMV}
\begin{eqnarray}\label{geneB}
H_{\tilde{\xi}}\big[t\big|\{B_m\}\big]&:=&
\sum_{m=1}^{\infty}\left\{\left[m+
\tilde{\rho}_m(t)\right]
\left[b^*_mb_m+b^*_{-m}b_{-m}\right]\right.
\nonumber\\ &+&\left. \tilde{\rho}_m(t)
\left[b^*_mb^*_{-m}+b_{m}b_{-m}\right]\right\},\\
\label{rhotil} \tilde{\rho}_m(t)&=&\frac{1}{2m}\left[
\frac{1}{4t^2}-\dot{g}(t)\right].\end{eqnarray}

On the other hand,we remember that, from our
discussion in Subsec. \ref{compleuni}, the variables
$\{\tilde{B}_m(t)\}$ corresponding to the canonical
pair $(\tilde{\xi},P_{\tilde{\xi}})$ are related to
$\{B_m(t)\}$ by the matrices $C_m(t)$ obtained from
Eq. (\ref{2}) with $f(t)=1$, namely
$B_m(t)=C^{-1}_m(t)\tilde{B}_m(t)$ $\forall m\in
\mathbb{N}$. In terms of $\{\tilde{B}_m(t)\}$ we then
get
\begin{equation}
H_{\tilde{\xi}}=H_{\tilde{\xi}}^0+
H_{\tilde{\xi}}\big[t\big|\{C^{-1}_m(t)\tilde{B}_m(t)\}
\big].
\end{equation}
Therefore, in the Schr\"odinger picture and obviating
the contribution of the zero modes (which are a
finite-dimensional system), the generator of the
dynamics of the fieldlike variables
$(\tilde{\xi},P_{\tilde{\xi}})$ is
$:H_{\tilde{\xi}}\big[t\big|\{C^{-1}_m(t)\hat{B}_m
\}]:+D$, where we have used
$\tilde{B}_m(t_0)=B_m(t_0):=B_m$, $\hat{B}_m$ denotes
the operator counterpart of $B_m$ obtained with the
complex structure $J$ (for simplicity, we will obviate
the use of a more accurate notation such as
$\hat{B}_m^J$ that would make explicit this fact), the
dots denote normal ordering with respect to $J$, and
$D$ is a c-number representing a possible zero-point
energy.

As we have commented in Subsec. \ref{compleuni}, any
invariant complex structure is related with $J_0$ by
means of a time-independent symplectic transformation,
$J=K_JJ_0K_J^{-1}$, with $K_J$ given in Eq.
(\ref{nn1}) \cite{CCMV}. Furthermore, we have shown in
Subsec. \ref{equivrepre} that, if the dynamics of the
canonical pair $(\tilde{\xi},P_{\tilde{\xi}})$ is
unitarily implementable with respect to $J$, then
$K_J$ admits a unitary implementation in the quantum
representation determined by the complex structure
$J_0$. Therefore, there exist unitary operators
$\hat{K}_J$ such that
\begin{equation}\hat{K}_J
\hat{B}_m\hat{K}_J^{-1}=({\cal K}_J)_m^{-1}
\hat{B}_m:=\hat{A}_m.\end{equation} In the
corresponding basis $\{A_m\}$, with
\begin{equation}
({\cal K}_J)_m^{-1}
B_m:=A_m:=(a_m,a^*_{-m},a_{-m},a^*_m)^T,
\end{equation} the complex structure $J$ has the same
matrix form as $J_0$ in the original basis $\{B_m\}$.
In other words, $J$ is block diagonal in terms of
$\{A_m\}$, with $4\times 4$ blocks equal to
$(J)_m={\rm diag}(i,-i,i,-i)$. The vacuum $|0>_J$
associated with the complex structure $J$ is simply
the state annihilated by the operators $\hat{a}_m$ and
$\hat{a}_{-m}$ $\forall m\in\mathbb{N}$. In total, we
arrive at the following expression for the quantum
generator of the dynamics of the canonical pair
$(\tilde{\xi},P_{\tilde{\xi}})$ (modulo the
contribution of the zero modes):
\begin{equation}\label{Deq}
\hat{H}_{\tilde{\xi}}(t):=\;
:H_{\tilde{\xi}}\big[t\big|\{C^{-1}_m(t)({\cal
K}_J)_m\hat{A}_m \}\big]:+D,\end{equation} where the
normal ordering is that corresponding to the
annihilation and creation operators $\{\hat{A}_m\}$.

A straightforward calculation shows then that
\begin{equation}\label{nor}
||\hat{H}_{\tilde{\xi}}(t)|0>_J||^2=|D|^2+
\sum_{m=1}^{\infty} |\gamma_m(t)|^2,\end{equation}
where
\begin{eqnarray}
\gamma_m(t)&=&2m
\kappa_m(t)\lambda_m^{*}(t)+\tilde{\rho}_m(t)
[\kappa_m(t)+\lambda_m^{*}(t)]^2,\nonumber \\
\nonumber
\kappa_m(t)&=&\kappa_m-i(\kappa_m+\lambda_m^{*})
\frac{g(t)}{2m},\\
\label{lambda} \lambda_m(t)&=&
\lambda_m-i(\kappa_m+\lambda_m)\frac{g(t)}{2m}.
\end{eqnarray}
Here $\{\kappa_m=\sqrt{1+|\lambda_m|^2}\}$ (which are
real) and $\{\lambda_m\}$ are the time-in\-dependent
coefficients of the symplectic transformation $K_J$
[see Eq. (\ref{nn1})]. We also remember that the
sequence $\{\lambda_m\}$ is square summable, because
$K_J$ is unitarily implementable with respect to
$J_0$.

We see from Eq. (\ref{nor}) that, for $|0>_J$ to
belong to the domain of $\hat{H}_{\tilde{\xi}}(t)$ at
any positive value of $t$, the sequence
$\{\gamma_m(t)\}$ must be square summable $\forall
t>0$. It then follows that $2m\lambda_m^{*}(t)$ must
be negligible compared with $1/\sqrt{m}$ when $m\to
\infty$ $\forall t>0$, because this factor is either
of order $1/m$ (i.e., its product by $m$ is bounded)
or it gives the leading term in $\gamma_m(t)$, which
has to be square summable. But this implies that
$\lambda_m(t)$ must be negligible compared with
$1/m^{3/2}$ $\forall t>0$. Since the time-dependent
part of $\lambda_m(t)$ is
$-i(\kappa_m+\lambda_m)g(t)/(2m)$, which is of order
$g(t)/m$, it is necessary that $g(t)$ be constant, so
that this contribution can be compensated by the
time-independent part. Therefore, we conclude that
$g(t)=g(t_0)=0$.

This singles out the momentum
$P_{\tilde{\xi}}=P_{\xi}$ of the CCMV formulation. In
the case of the CCMV canonical pair, our condition on
the vacuum is satisfied with the choice of complex
structure $J_0$ [assuming $|D|< \infty$ in Eq.
(\ref{Deq})]. In order to see this note that, with
$K_J$ being the identity and $g(t)=0$, one obtains
$\gamma_m(t)=1/(8mt^2)$, which is clearly square
summable $\forall t>0$.


\begin{thebibliography}{99}

\bibitem{misn} C.W. Misner, Phys. Rev. D {\bf 8},
3271 (1973).

\bibitem{berger} B.K. Berger, Ann. Phys. (N.Y.)
{\bf 83}, 458 (1974); Phys. Rev. D {\bf 11}, 2770
(1975); Ann. Phys. (N.Y.) {\bf 156}, 155 (1984).

\bibitem{hs} V. Husain and L. Smolin, Nucl. Phys.
{\bf B327}, 205 (1989).

\bibitem{guillermo} G.A. Mena Marug\'an, Phys. Rev. D
{\bf{56}}, 908 (1997).

\bibitem{pierri} M. Pierri, Int. J. Mod. Phys. D {\bf
11}, 135 (2002).

\bibitem{ccq-t3} A. Corichi, J. Cortez, and H. Quevedo,
Int. J. Mod. Phys. D {\bf 11}, 1451 (2002).

\bibitem{torre-prd} C.G. Torre, Phys. Rev. D {\bf 66},
084017 (2002).

\bibitem{come} J. Cortez and G.A. Mena Marug\'an,
Phys. Rev. D {\bf 72}, 064020 (2005).

\bibitem{cocome2} A. Corichi, J. Cortez, and G.A.
Mena Marug\'an, Phys. Rev. D {\bf 73}, 041502 (2006).

\bibitem{cocome} A. Corichi, J. Cortez, and G.A. Mena
Marug\'an, Phys. Rev. D {\bf 73}, 084020 (2006).

\bibitem{CCMV} A. Corichi, J. Cortez, G.A. Mena
Marug\'an, and J.M. Velhinho, Classical Quantum
Gravity {\bf 23}, 6301 (2006).

\bibitem{gowdy} R.H. Gowdy, Phys. Rev. Lett. {\bf 27},
826 (1971); Ann. Phys. (N.Y.) {\bf 83}, 203 (1974).

\bibitem{ash-mag} A. Ashtekar and A. Magnon, Proc.
R. Soc. Lond. A {\bf 346} 375, (1975); A. Ashtekar and
A. Magnon-Ashtekar, Pramana {\bf 15}, 107 (1980).

\bibitem{poincare} A. Corichi, J. Cortez, and H.
Quevedo, Ann. Phys. {\bf 313}, 446 (2004).

\bibitem{wald} R.M. Wald, {\it Quantum Field Theory in
Curved Spacetime and Black Hole Thermodynamics}
(Chicago Press, Chicago, 1994).

\bibitem{polymer}
T. Thiemann, Classical Quantum Gravity {\bf 15}, 1487
(1998); A. Ashtekar, J. Lewandowski, and H. Sahlmann,
Classical Quantum Gravity {\bf 20}, L11 (2003); A.
Ashtekar and J. Lewandowski, Classical Quantum Gravity
{\bf 21}, R53 (2004); W. Kaminski, J. Lewandowski, and
M. Bobienski, Classical Quantum Gravity {\bf 23}, 2761
(2006).

\bibitem{note0} We employ a system of units with
$c=4G/\pi=1$, $c$ and $G$ being the speed of light and
Newton's constant.

\bibitem{noteZ2} One could also
consider translations of the field by a homogeneous
solution to the associated Klein-Gordon equation, but
this would only affect the zero mode of the field and
not an infinite number of degrees of freedom. Hence,
this kind of redefinition is irrelevant when
considering quantum representations.

\bibitem{note} Actually, one could cope with this
coupling at the cost of considering complex structures
that depend on $P$.

\bibitem{Z5} The use of invariant
complex structures is common to previous works on the
Gowdy model, and there is no motivation to generalize
this procedure when looking for a quantization of the
constraint $C_0$, given that a large class of
invariant complex structures exists. Moreover, unitary
implementation of classical symmetries in a natural
invariant way, whenever available, is a standard
feature of quantum field theory.

\bibitem{abra} {\it Handbook of Mathematical
Functions}, edited by M. Abra\-mowitz and I.A. Stegun,
NBS Appl. Math. Ser. --No. 55 (U.S., GPO, Washington,
DC, 1970), 9th ed.

\bibitem{fock}
Y. Choquet-Bruhat, C. DeWitt-Morette, and M.
Dillard-Bleick, {\it Analysis, Manifolds and Physics,
Part I: Basics} (Elsevier, Amsterdam, 1996); B.S. Kay,
Commun. Math. Phys. {\bf 62}, 55 (1978).

\bibitem{noteZ1} For a given complex structure
$J$, we use the expression ``Fock vacuum'', or simply
``vacuum'', to  refer to the standard unit vector of
the corresponding Fock space that belongs to the
kernel of all the annihilation operators defined by
$J$.

\bibitem{un1}  D. Shale, Trans. Am. Math. Soc.
{\bf 103}, 149 (1962); R. Honegger and A. Rieckers, J.
Math. Phys. {\bf 37}, 4292 (1996).

\bibitem{ash-mag2} A. Ashtekar and A. Magnon-Ashtekar,
Gen. Rel. Grav. {\bf 12}, 205 (1980).

\bibitem{torre2} C.G. Torre, Classical Quantum Gravity
{\bf 23}, 1543 (2006).

\bibitem{gold} See, e.g, H. Goldstein,
{\it Classical Mechanics} (Addison-Wesley, Rea\-ding,
MA, 1980), 2nd ed.

\bibitem{noteZ3} Clearly, $f(t)$ and $g(t)$
have the same regularity properties as $F(t)$ and
$G(t)$, and $f(t)$ vanishes nowhere. Explicitly,
$f(t)=F(t)/F(t_0)$ and $g(t)=F(t_0)G(t)-G(t_0)F(t)$.

\bibitem{nota} Actually, the expression for unitary
implementability of the evolution between arbitrary
times $t'$ and $t$ treated as a symplectic
transformation on our phase space is that ${\tilde
U}^{-1}(t',t_0){\tilde U}(t,t'){\tilde
U}(t',t_0)J_{t_0}{\tilde U}^{-1}(t',t_0){\tilde
U}^{-1}(t,t'){\tilde U}(t',t_0)$ be equivalent to
$J_{t_0}$ $\forall t, t'>0$. This condition is clearly
equivalent to Eq. (\ref{u7}).

\bibitem{noteZ4} The same conclusion
can be obtained directly from the discussion presented
in Sec. \ref{nogo} if one starts with transformation
(\ref{1}). In that case, Eq. (\ref{3}) must be
replaced with $\tilde {\cal U}_m(t,t_0) = C_m(t){\cal
U}_m(t,t_0)C^{-1}_m(t_0)$ because $C_m(t_0)\neq {\bf
1}$ now. This affects some of the formulas given in
Sec. \ref{nogo} but does not modify the conclusion.

\bibitem{mizo} For a geometrical interpretation of
this homogeneous variable in terms of the central
element of the Geroch group, see A. Mizoguchi, Phys.
Rev. D {\bf 51}, 6788 (1995).

\bibitem{kolfo} See, e.g.,
M. Reed and B. Simon, {\it Methods of Modern
Mathematical Physics\/ I: Functional Analysis}
(Academic Press, San Diego, 1980).

\end{thebibliography}
\end{document}